# The Cross-Survey Decade: A Call to Action

## Toward cross-survey astrophysics

Gioia Rau[1,*], Robert Benjamin[2], Federica Bianco[3], Ranga-Ram Chary[4], Andy Connolly[5], Cecilia Garraffo[6], Suvi Gezari[7], Leanne P. Guy[8,9], Željko Ivezić[5], Stephanie Juneau[9], Vicky Kalogera[10], Mansi M. Kasliwal[11], François Lanusse[12], Zack Li[13], Rachel Mandelbaum[14], Peter Melchior[15], Stella Offner[16], Antonella Palmese[14], Jason Rhodes[17], Edward Schlafly[18], Kartik Sheth[19], Rachel Street[20], Michael Troxel[21], Tony Tyson[22], Beth Willman[23], Michael Wood-Vasey[24], Yuanyuan Zhang[9]

[1]Renaissance Philanthropy
[2]University of Wisconsin
[3]University of Delaware
[4]Caltech/IPAC, UCLA
[5]University of Washington
[6]Center for Astrophysics | Harvard & Smithsonian
[7]University of Maryland
[8]Vera C. Rubin Observatory
[9]NSF NOIRLab
[10]Northwestern University
[11]California Institute of Technology
[12]Université Paris-Saclay, Université Paris Cité, CEA, CNRS, AIM
[13]University of California, Berkeley
[14]Carnegie Mellon University
[15]Princeton University
[16]University of Texas at Austin
[17]Jet Propulsion Laboratory, California Institute of Technology
[18]Space Telescope Science Institute
[19]Aix-Marseille University, CNRS, CNES, LAM/Georgetown University
[20]Las Cumbres Observatory
[21]Duke University
[22]University of California, Davis
[23]LSST Discovery Alliance
[24]University of Pittsburgh

*_Corresponding author: gioiarau.space@gmail.com_

## Abstract

Astronomy has long benefited from coordinated multiwavelength observations across facilities. The turn toward integrated, cross-spectrum analysis was consolidating by the 1980s [57], the Great Observatories established the scientific power of the approach [56], and the current generation of wide-field surveys demonstrated it at scale.
By 2027, three flagship wide-field surveys will be operating simultaneously from ground and space, surveying overlapping sky, representing a combined US and European public investment exceeding $6 billion. These will produce overlapping petabyte-scale datasets over thousands of square degrees. That is many orders of magnitude beyond anything attempted, and a different class of challenge: the observations are no longer the bottleneck; realizing the joint scientific return is now an engineering and institutional challenge, and depends on shared computational infrastructure and coordination mechanisms.

Decades of community studies have delivered a clear verdict: combining these datasets doesn't merely improve precision. For ground-breaking science ranging from weak lensing to transient discovery and Galactic-plane astronomy, joint data processing and subsequent analysis unlocks each survey's full potential, achieving the highest accuracy constraints on critical physical parameters. Yet, because each mission is funded and executed independently, the collaborative infrastructure required -- such as joint pixel-level processing, cross-calibration and cross-validation, data access, and the personnel to build them -- currently falls outside any single mission or institution's mandate.

We issue a call to action for cross-survey science infrastructure: the deliberate production of joint-survey data products, processing pipelines, and validation frameworks as infrastructure is as essential to these surveys' science as the telescopes themselves. We propose four pillars: (1) joint pixel-level processing and validation; (2) an AI-ready data substrate for scientific foundation models; (3) standardized, interoperable data access across surveys, democratizing participation in astrophysical discovery; and (4) dedicated personnel and career pathways. We outline concrete steps for policymakers, agencies, observatories, universities, the research community, and philanthropy; and we argue that the moment to act is now: while foundational technical choices can still be aligned at a fraction of the cost of reconciling them later.

## 1. Motivation and Background

The NSF–DOE Vera C. Rubin Observatory began its ten-year Legacy Survey of Space and Time (LSST) in June 2026 [1]; and NASA's Nancy Grace Roman Space Telescope is set to launch later this year (no earlier than August 30, 2026). While this Call to Action anchors on the Rubin-Roman-Euclid priority (beginning with the data already in hand from Rubin and Euclid, which is what prepares the deeper Rubin-Roman combination) the thinking behind it establishes an extensible model that expands to a broader network. The same model scales first to the surveys nearest in kind - SPHEREx, DESI, UVEX and Gaia - and then across the spectrum, to radio and sub-millimetre facilities such as SKA and ALMA, to high-energy missions such as NewAthena, and to the archival holdings from Spitzer, Chandra and their contemporaries, which become newly valuable once a joint framework exists. It extends equally to the new public and private wide-field facilities coming online over the next few years.

The scientific value of synergistically combining observations across wavelengths has been recognized for decades, and repeatedly acknowledged in the literature: Roman's time-allocation report [2] highlights the necessity of Rubin photometry for its flagship cosmology program, and Roman's Deep Tier is sited within Rubin's Deep Drilling Fields; dedicated working groups have produced recommendations on Rubin-Euclid derived data products [5], on Rubin-Euclid scientific synergies [6], and on joint Roman-Euclid microlensing science [7]. The analysis has been done. What has never followed is the infrastructure to act on it.

Each survey pushes technological advances to produce new scientific insights and delivers on the requirements it was designed and funded to meet. Rubin's data ecosystem is built for community access at unprecedented scale; Roman's launch caps decades of engineering excellence and exceeds its technical requirements in many key aspects of the observatory; Euclid is already delivering space-resolution imaging over thousands of square degrees. Each project's teams are preparing to produce exactly what they were scoped, funded, and asked to produce.
But those designs predate the full scale of the scientific opportunity now emerging at the intersection of these datasets. Responsibility for each mission rests, appropriately, with its sponsoring agency or agencies; responsibility for what becomes possible only when the data are combined rests, at present, with no single mission or agency.

This shift has a long arc behind it. From the 1960s through the 1980s, multiwavelength astronomy established that understanding an object required observing it across the spectrum rather than through a single window [57]. From the 1990s through the 2010s, that scientific concept became a coordination problem addressed through the Great Observatories, deep-field campaigns, and community efforts toward interoperable data access. What the rest of 2020s and 2030s and beyond require is different in kind: not coordinating observations across facilities, but processing petabyte-scale surveys jointly, at the pixel level, rather than combining their catalogues after the fact.

The distinction matters. In the early 2000s the Great Observatories Origins Deep Survey (GOODS) [3] pioneered forced photometry, using high-resolution HST imaging as a positional and morphological prior to extract consistent photometry from Spitzer's far lower-resolution pixels, reducing photometric scatter and catastrophic redshift outliers relative to simple catalogue matching. Even at a mere 330 arcmin$^2$, this was an extraordinary computational and human undertaking. What would fully integrate Rubin, Euclid, and Roman goes further still: genuine simultaneous joint pixel-level processing, not one survey's resolution merely informing extraction from the other -- at a scale, thousands of square degrees, many orders of magnitude larger. That is not an incremental extension of what GOODS did, but rather a different class of challenge; and it is why the bottleneck has moved from the telescopes to the infrastructure between them.

The moment to act is now: the computational burden that once made joint pixel-level processing prohibitive is no longer the barrier it was a decade ago. More urgently, these surveys are early enough in their lifecycles that foundational technical choices -- coadd projections, for example -- can still be aligned; doing so now is far cheaper than reconciling a divergence later, and that window will close in the near term.
Without immediate intervention, massive scientific discovery will be left on the table.

## 2. Science Unlocked by Cross-Survey Coordination

The foundational scientific synergy between these flagship facilities has been described across the literature and extensively documented in comprehensive community reports, including reports on Rubin-Roman synergies [4], Rubin-Euclid derived data products recommendations [5], Rubin-Euclid scientific synergies [6], and joint Roman-Euclid microlensing science [7] -- and the case for the joint pixel-level processing capability itself, requested of the Astro2020 decadal survey in 2019 [8]. What the literature establishes is clear: combining surveys is what unlocks the full potential of each survey, for many of the science cases -- as the legacy multiwavelength surveys demonstrated, with COSMOS [60] alone yielding results from the distribution of dark matter to the nature of reionization.

Improved science from combining these existing surveys -- and, as the field matures, from expanding to future surveys and wavelengths -- comes from three complementary strengths. The first two benefit most science, the third is more specialized: (a) wavelength complementarity: Rubin's broadband optical paired with Euclid and Roman's broadband and spectroscopic near-infrared sharpens photometric redshifts and object modeling and disentangles the effects of dust and environment; (b) space-based resolution: Roman and Euclid deblend objects that appear confused or merged in Rubin's images -- a problem affecting more than half of all sources at full Rubin depth (e.g., [64]) -- with Euclid delivering wide-field space resolution now and Roman going deeper over a smaller area; (c) wide-baseline parallax: together, Rubin and Roman deliver optical plus near-IR astrometry beyond what any ground-based pair could achieve -- mainly for solar-system and near-Milky-Way science, such as identifying Earth-impacting asteroids.

Some science cases needing only photometry, lightcurves, or cutouts of matched objects can be served by the planned public data products or straightforward extensions to them, such as forced photometry in one survey based on detections in the other. Others -- weak lensing chief among them -- require joint pixel-level processing to fully reap the benefits of imaging from multiple surveys: space-resolution priors identify which Rubin sources are blended, and deblended photometry measured through consistent apertures across surveys yields the consistent colors, and therefore the accurate photometric redshifts, that unbiased redshift tomography for w(z) depends on.
This processing can be done whenever the data are available, with no change to either survey's observing strategy (though decisions taken now, like common coadd projections, would make it substantially simpler and cheaper later) and it yields as a by-product the joint, deblended multi-wavelength photometry, morphology, and photometric redshifts that benefit nearly every other science case.

We summarize below the science cases that reach their full potential only through cross-survey coordination; Appendix A presents the complete landscape in detail.
**Cosmology.** Rubin delivers weak-lensing shapes over ~18,000 square degrees, but its photometric redshifts suffer catastrophic outliers exactly where it matters, and ground-based blending biases the shear. Roman's near-infrared photometry significantly improves the photo-z problem, and its stable space-based PSF offers a route to calibrating many critical ground-based systematics at the pixel level; jointly, the two tighten dark-energy equation-of-state constraints beyond what either achieves alone and cross-check systematics neither can diagnose alone. This is the flagship case that cannot be realized from the planned public products: it requires joint pixel-level detection and modeling to reach maximum depth, deblend sources, and obtain consistent photometry, morphologies, and shape measurements across the two surveys. These joint measurements reduce photometric-redshift outliers and enable more robust calibration of shear and selection effects. The same processing yields the host-galaxy models that sharpen supernova cosmology and supports nearly every science case below.

**Time-domain and transients.** The single-survey time-domain foundations are rapidly maturing on both sides. On the Roman side, RAPID (Roman Alerts Promptly from Image Differencing; [10]) -- a Caltech/IPAC Project Infrastructure Team -- will deliver prompt image differencing, a public alert stream of transient and variable candidates, and forced-photometry services, with machine-learning classification feeding the community brokers. On the Rubin side, the Legacy Survey of Space and Time (LSST) alert stream and community brokers are already operating and beginning to ingest multi-messenger streams. Each survey's transient infrastructure is being built to do precisely what it was chartered to do -- and each charter is, deliberately, single-survey. What remains outside either survey's scope is the joint layer: brokers that merge the Rubin and Roman streams and enrich each alert with joint metadata, triggering jointly on events that require both surveys, and modeling the shared sky as one scene. The stakes are not incremental: joint searches can identify events that either survey alone would miss or classify ambiguously. For high-redshift supernovae, for example, deep Rubin non-detections can help distinguish Roman-detected events from lower-redshift contaminants. Joint identification can also enable rapid, targeted follow-up: sources of high scientific interest can be flagged and passed to other facilities while the event is still observable, which for the shortest-lived classes is a matter of hours.

**Galaxies, AGN, and black holes.** The pixel-level products required for cosmology also deliver joint optical+near-IR depth and deblending, which open new reach on dwarf and low-surface-brightness galaxies; those are among the hardest objects for either survey alone to identify and weigh. The resolution of the space-based data cleanly separates AGN light from starlight, bearing on the origin of supermassive black holes, and the same joint imaging enables dust maps for the majority of all galaxies. Individual teams can and do assemble joint measurements for particular samples, but doing so per-application means repeating pixel-level work at full cost and arriving at selection functions that cannot be compared across studies. If built once, then the substrate delivers this science at population scale with a characterized selection function.

**Stars and the Milky Way.** At high Galactic latitude, independent Roman and Rubin catalogs can reasonably be cross-matched. However, in the Galactic plane, where most of the Galaxy's stars and dust reside, extreme crowding and differential extinction between the optical and NIR bands makes catalog cross-matching ambiguous and compromises the underlying crowded Rubin photometry. Taking full advantage of the Rubin optical photometry requires joint photometry from joint pixel-level processing. This matters acutely for the Roman Galactic Plane Survey [11], which primarily observes at F129 and redward, leaving a limited lever arm for constraining stellar types, which becomes more valuable when complemented with Rubin's optical photometry. Joint pixel processing allows the combination of both surveys to massively extend the map of Galactic structure and kinematics beyond Gaia's reach in crowded regions.

**Solar system.** Joint parallax of moving objects -- leveraging Roman's wide field and non-destructive reads with Rubin's cadence -- delivers astrometry beyond any ground-based pair, with direct application to detecting and characterizing potentially Earth-impacting asteroids and to probing the outer solar system.

**Microlensing**. Microlensing events are transient and unrepeatable, so the parallax baseline between simultaneous observers is the measurement. Roman and Euclid observing the Galactic bulge together provide a space-based baseline that resolves the lens mass and distance degeneracy directly, enabling mass measurements of free-floating planets and, in favourable cases, exomoons [7]; adding Rubin's optical cadence from the ground extends the wavelength lever arm and the temporal coverage on the same events. None of this can be recovered afterwards from independent catalogues; the observations have to be processed together.

## 3. Call to Action

Historically, the evolution of astronomical surveys has been toward progressively wider, deeper, and faster single-survey instruments. Each generation of facilities has allowed astronomers to see further and larger portions of the sky, more often, to greater depth than the one before - from targeted pointed observations to the wide-field, high-cadence surveys of the current decade [12, 13].

Today's flagship surveys are rooted in scientific priorities and planning efforts extending back decades - a testament to sustained foresight, and a reminder that no plan conceived at that time could have anticipated all that has since become possible [13, 14, 15]. The pace of technological and scientific progress has made something clear: these surveys reach their full scientific value only when their data are combined [4, 6, 16, 17, 51]. This calls for a corresponding evolution in how surveys are supported and operated: from optimizing each survey's delivery in isolation, toward treating the joint layer between them as infrastructure in its own right [16, 19]. The window for this is open now, and it is the cheap window: foundational choices (e.g., data and metadata formats, coadd projections, processing conventions) can still be aligned across the surveys by design [55], at a fraction of the cost of reconciling divergent choices after the fact (that would mean reprocessing at survey scale, rather than agreeing on conventions once).

Leaving this to each project individually runs into a structural limit. The natural approach is for one survey to ingest another's data into its own pipeline -- but each survey's data is best understood by the team that built its pipeline, and a project processing another's data works without that expertise, recreating the same gap under a different name [16, 20]. Instrumental artifacts make this concrete: persistence, cosmic-ray afterglows, snowballs, stray light. Each team knows its own, and no external group can reconstruct that knowledge reliably. Consistent flagging across surveys -- common bit conventions, and tools trained on each survey's own known artefacts -- therefore requires the expertise to be pooled.

The community is already demonstrating demand for this capability [4,19]. In the absence of a standardized option, scientists have adapted single-survey pipelines and developed custom cross-matches to support joint analyses -- resourceful efforts that demonstrate both the scientific appetite and the feasibility of cross-survey work. IRSA's tutorials for retrieving matched cutouts from simulated Roman–Rubin fields are a case in point: it works, and it works by hand, one dataset at a time [41]. But these efforts often require teams to develop overlapping bespoke machinery, while combinations lacking harmonized calibration and validation can introduce systematics that are difficult to distinguish from genuine signals. The resulting products can also be difficult to sustain, reproduce, or extend as additional surveys come online or existing data products and formats evolve. The demand is established; what remains is to provide the shared infrastructure that can support it efficiently and at scale.

Agencies and projects understandably prioritize delivery of products satisfying each survey's defined requirements and established science cases before committing resources to activities beyond that scope. Yet demonstrating the full value of joint science requires precisely the cross-survey work that is not consistently supported within existing project structures, and producing that evidence can be difficult without dedicated funding for cross-survey processing and analysis. This is the natural equilibrium of a funding architecture organized primarily around individual facilities and single missions. Work that draws comparably on datasets, expertise, and infrastructure supported by different agencies may fit imperfectly within any one program's remit or review criteria. Research that depends equally on datasets supported by multiple agencies may not align neatly with any single program's criteria, leaving cross-survey science in an area of shared interest but diffuse ownership.

In this community perspective article, we present a call to action for cross-survey science infrastructure: the deliberate treatment of joint data products, joint processing, and joint validation as infrastructure in their own right, as essential to realizing the science of these surveys as the telescopes themselves. Building this requires four things: (1) joint pixel-level processing and validation; (2) an AI-ready, cross-survey data substrate; (3) standardized, interoperable access across surveys; and (4) people and career pathways dedicated to designing, building and sustaining it.

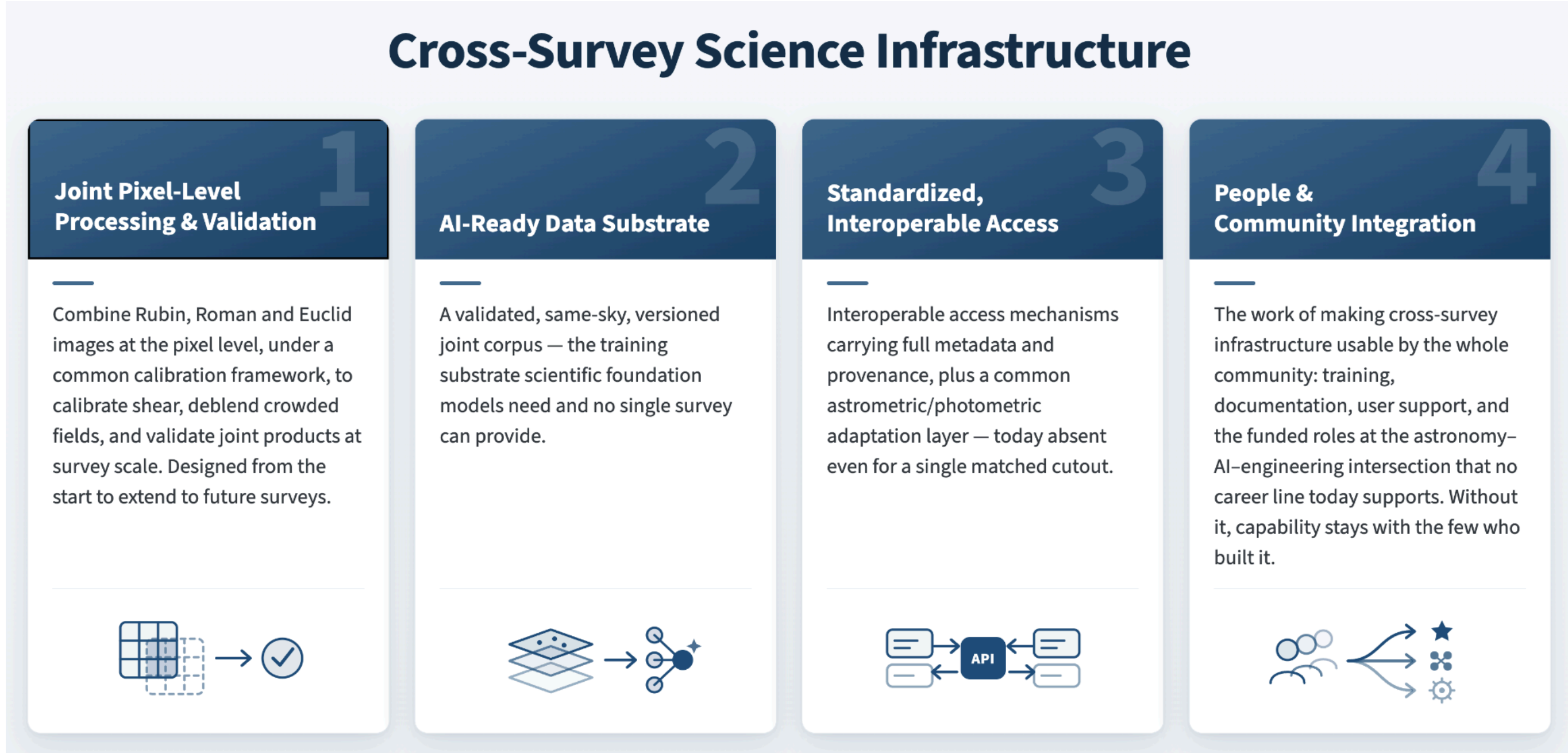

**Figure 1**: The four pillars of cross-survey science infrastructure. Each is specified by community studies and demonstrated in prototype; none is currently funded or owned as a shared capability.

### 3.1 Joint Pixel-Level Processing and Validation

Combining Rubin and Roman images at the pixel level, not just their derived catalogs, is necessary to calibrate shear and selection for weak lensing, deblend crowded fields, and produce accurate photometry that neither survey can deliver alone. The algorithms exist and have been demonstrated in isolated cases (e.g., [16, 21, 52, 53]), but they have not been tested at survey scale, and -- critically -- no shared tools exist to validate their output across surveys at scale. A joint shear catalog may show a bias absent from either input catalog; nothing currently tells us where such a bias comes from, or how to calibrate results at a depth exceeding either survey's own internal calibration. Closing this requires a common calibration framework spanning astrometry, photometry, PSF characterization, and noise modeling across both surveys -- not two internally-consistent calibrations that have never been reconciled with each other [e.g., 22[1]]. The DES-KiDS joint analysis is instructive: the first combined catalogue-level analysis between two weak-lensing collaborations, it ruled uniform systematics calibration out of scope even with both teams involved [61]; not for lack of expertise, but because the work sat outside what either collaboration was scoped to do.

For datasets at the precision Roman and Rubin jointly promise, this likely requires inference on a shared underlying pixel or latent space [23, 24], rather than combining independently-modeled noise characterizations after the fact. Validation must also extend to systematics flagging -- ghosts, stray light, diffraction spikes, bad pixels -- applied consistently across surveys, with common bit-flag conventions and PSF kernels reconciled epoch to epoch wherever space-based and ground-based data are matched.

Investment has historically concentrated on the hardware and instruments that produce these data; the software infrastructure to mine the results fully is the natural next layer. The precision these surveys jointly promise raises the bar for validation: subtle systematics introduced when combining independently processed data can be indistinguishable from genuine physical signals. Uniform joint-processing pipelines, paired with the joint simulations needed to validate them [25, 41], are what allow the community to tell the difference - and to place full confidence in the discoveries that follow. Shared, same-sky simulations are essential for validating joint analyses: a common astrophysical sky must be propagated through each survey's instrument, noise, and selection effects. This is also what makes arbitrary pairs of datasets combinable reliably, as it has in CMB cosmology for two decades. OpenUniverse2024 [41] is the closest existing prototype, with matched Rubin-Roman simulations produced through a third-party effort that also demonstrates such coordination is achievable at scale; a full Rubin-Roman-Euclid simulation suite does not yet exist. The capability should be designed from the start to extend beyond the founding pair: SPHEREx, UVEX, and future surveys should be able to join the same framework rather than requiring new bilateral reconciliations.

No funded programme exists to do this on real data. Simulated joint products have been produced [41], and the algorithms have been demonstrated in isolated cases, but nothing is scoped or resourced to carry it out at survey scale when the data arrive. A concrete first step could be a joint pixel-level processing demonstration: Roman and Rubin stamps processed jointly in a shared deep field such as the Euclid Deep Field South (EDF-S), where all three surveys overlap [26].

---

[1]Longley et al. perform a unified catalogue-level reanalysis across DES, HSC and KiDS, applying consistent modelling choices to each; but do not reconcile the surveys' calibrations against one another. That step, at catalogue level and still more at the pixel level, remains undone.

### 3.2 An AI-Ready Data Substrate

The most valuable joint product for the machine-learning community is a training substrate (a large, uniformly processed dataset that models learn from) for scientific AI. A high-quality cross-survey training substrate requires reliable source association and well-characterized astrometric and photometric calibration, with uncertainties and provenance preserved in the data products (and not reconstructed downstream) [51]. AstroCLIP [50], which learns jointly from images and spectra of the same galaxies, illustrates the value of such multimodal training. These requirements echo the AI+MPS community's call for domain-informed, multimodal foundation models supported by scalable scientific data infrastructure, rigorous uncertainty quantification, and data provenance [27]. Rubin, Roman, and Euclid could jointly provide such a corpus: billions of sources spanning optical and near-infrared wavelengths, time-domain coverage, and complementary spatial resolutions over common fields. No existing project yet is scoped to produce this.

Rubin will detect vast numbers of sources that no catalogue can classify. Most of what appears unusual in such data is an artifact of the instrument or the observing conditions rather than of the sky. Models trained to separate instrumental effects from physical signals [e.g., 54] can tell the difference -- turning an unmanageable flood of unclassified detections into a tractable search for objects that are new. Building these models as open resources from the start is what lets the community build on and improve them, and extends naturally to prioritizing models that help triage the resulting flood of candidates toward the ones worth following up.

Astronomy is unusually well placed to supply this: the field standardized its data format decades ago, invested in open community tooling, and produces data free of the privacy and commercial constraints that limit AI development elsewhere -- a combination that makes it a unique environment for building and testing scientific AI models [59].

The joint data products of Section 3.1 have to be usable. Each survey provides an excellent platform for its own data -- the Rubin Science Platform and Roman's Nexus among them -- but each serves one survey's holdings, in that survey's formats and conventions. Working across surveys means working across those boundaries.

The reason this substrate doesn't exist yet is not scientific but architectural: there is no front door. A coherent, cross-survey interface -- served from validated, deblended, multi-wavelength catalogs -- would not only enable the AI use case above; it would democratize access to a multi-petabyte scientific dataset that today only well-resourced teams can practically use. Separate archives, separate formats, no common identifier, no benchmark, no clean Application Programming Interface (API) -- even the joint products the community has carefully specified on paper remain unbuilt. That layer is shared infrastructure: everyone needs it, and it serves no one to build several versions of it.

The standardization half of this challenge has already been demonstrated. The Multimodal Universe [28] (an open, approximately 100 TB dataset assembled by a broad collaboration and hosted at the Flatiron Institute) brings images, spectra, and time series from many surveys into a common schema with benchmark tasks. It shows that the pattern works: when a curated, benchmarked corpus becomes available, a modeling community can rapidly form around it, as happened with ImageNet [29, 30]. But standardizing existing survey products is not the same as creating a jointly processed, same-sky dataset. As the Multimodal Universe itself notes, its cross-matched samples remain limited by the depths and footprints of the underlying surveys, and by the extent to which those footprints overlap. Deep Rubin-Roman joint processing would create a qualitatively different substrate: coextensive measurements, joint source models, and validated cross-survey associations and labels derived from the same sky.

The astronomy-AI ecosystem is beginning to take shape through efforts such as Polymathic AI [31] and the NSF-Simons AI Institutes SkAI [32] and CosmicAI [33], each working downstream of data products generated by individual surveys. The critical missing piece is upstream: a large, validated, same-sky Rubin-Roman corpus on which multimodal models can be trained, tested, and benchmarked, with Euclid extending its wide-area near-infrared coverage and the same architecture generalizing to the surveys and wavelengths that follow. The payoff extends beyond astronomy: a physically grounded multimodal dataset at this scale would provide a template for other data-rich sciences. AlphaFold's breakthrough in protein-structure prediction depended on nearly fifty years of experimentally determined structures deposited in a single, standardized, openly accessible repository [58]. The corpus came first; the models followed; astronomy has yet to build its equivalent.

### 3.3 Standardized, Interoperable Access Across Surveys

Access is the bottleneck beneath both Sections 3.1 and 3.2. International Virtual Observatory Alliance (IVOA) standards [37] make individual archives programmatically queryable, but no common layer today serves harmonized Rubin, Roman, and Euclid data products: obtaining even a single matched cutout means authenticating against separate archives, reconciling coordinate conventions, and hand-building the cross-match and its validation. This is not a storage problem: physical co-location of the archives is neither realistic nor necessary. Roman, Rubin, and Euclid data can instead be exposed in interoperable cloud regions.

Making the data reachable is not the same as making it usable for everyone: shared, low-cost compute -- the Open Science Grid, or facilities like NOIRLab's Astro Data Lab [42, 43] and ESA Datalabs [34] -- allows researchers at smaller institutions to analyze data across interoperable cloud environments without prohibitive computing costs.

This kind of access is achievable: HATS/LSDB [35] (a catalog format and analysis framework developed by the LINCC Frameworks team [36] and its collaborators, in which catalogs are partitioned by position on the sky so they can be queried and cross-matched efficiently) already works on billion-source catalogs and now supports Rubin Data Preview 1. The near-term priority is to make Roman catalogs available through the same interoperable framework from the outset, rather than adding them later.

The same is true on the alert-stream side: recent community workshops[2] made clear that the existing community brokers are well positioned to provide cross-matched Rubin-Roman streams -- the open questions are design-level, not technical. As LSST and Zwicky Transient Facility (ZTF) alert streams are now being cross-matched in real time, this approach is already being tested on real survey data. Early prototypes show this is tractable: the community has already begun building programmatic interfaces that expose alert-stream information to automated systems [e.g., 68] -- an early instance of exactly this access layer emerging on its own.

[2] For example, at the June 2026 Rubin Broker Workshop, a dedicated session addressed writing Model Context Protocol (MCP) APIs so brokers could expose alert-stream information programmatically -- an early instance of exactly this access layer.

It is feasible to pull data for a given region of sky from each survey's own archive, process it jointly at any convenient computing center, and write the results back to whichever archive is best suited to hold them [e.g., 8]. The Roman High-Latitude Imaging Survey Project Infrastructure Team (HLIS PIT) [63] is already demonstrating this pattern with simulated Roman data: low-level data pulled from the cloud, processed at a university, higher-level products returned for archiving. What is missing are standardized interoperable access mechanisms[3] that carry the full metadata joint processing requires (background and weight maps, masks, PSF models, segmentation maps) while preserving provenance. Without this, every cross-survey effort re-invents the same access layer from scratch. Above that sits a second bottleneck: an adaptation layer to standardize astrometry and photometry across surveys, so that joint processing and joint training can draw on consistent inputs rather than each rebuilding the translation independently. This second layer is also critical for validating the results of joint processing.

A near-term, low-cost extension of the same idea: Roman's planned RAPID forced-photometry service could be made programmatically accessible to the LSST community brokers, so that a Rubin alert's position can be queried against Roman's photometric history in the overlap region and fed directly into transient classification; once available, Roman's prism and grism spectroscopy of the host galaxy -- served, as planned, through Roman's own archive -- would sharpen that classification further. The service is being built for Roman alone; the cross-survey interface is the small, unfunded step. Extending it to joint modeling from all imaging and spectroscopic data sources would further improve the physical understanding of transients.

### 3.4 People and Career Pathways

No established career path is dedicated to cross-survey infrastructure. And none of the above can be built without a sustained human infrastructure: people funded to do exactly this work. The talent this requires is inherently interdisciplinary, spanning astronomy, software engineering, and machine learning [49]. People who sit across these fields currently fall between every agency's mandate and every university/research lab incentive structures of the field, which have historically rewarded individual publications over tool-building work: creating shared tools the whole field relies on is currently not a wise career choice. Even the field's most successful software-infrastructure efforts are funded as fixed-term projects, with no continuation path -- so the strongest people at this intersection live appointment to appointment, by design. Career paths exist, but each sits inside a single facility or project; observatory centres sustain permanent software staff, and some project streams fund multi-year appointments. None spans across surveys or agencies. Universities compound it, rewarding publications over tool-building, so building shared infrastructure is rarely the career-advancing choice.

Astronomy has always recognized its instrument builders; the joint layer between surveys is the instrument of the cross-survey decade, and it deserves the same standing. What the pathway requires is known: funded roles whose explicit deliverable is the shared tooling itself, recognition

[3] For example building on the interoperability foundations the IVOA [37] has long established

and citation norms that make such work count, and stable, multi-year institutional homes -- the research-software-engineer model and survey-embedded fellowships already point the way[4]. Without them, a generation of exactly the people this effort needs will be lost to fields that do offer a path.

# 4. What to do

The development of new scientific capabilities is an interwoven process among many stakeholders, and this call addresses the research community first. This section identifies potential steps each can take.

### 4.1 Policymakers and Federal Agencies

- Make PI-level cross-agency grants viable. Work spanning multiple agencies is often split across separate panels, funding lines, and reporting structures, while the PI remains accountable to only one agency. Joint mechanisms would support genuine cross-agency expertise and allow ambitious projects to be planned and sustained as integrated efforts -- precisely the work that merits incentive: innovation no single agency could deliver alone. Agencies could partner to open robust funding mechanisms for cross-survey work; for example, joint proposal calls with clear accountability and decision processes, where one agency leads selection with input from the other, and reporting guidance follows the same structure.
- Remove the barriers that force single-survey defaults. Interagency dependencies have at times prevented missions from committing to products that rely on another agency's data[5].
- Fund the software and processing layer as strategic infrastructure, protected from diversion when budgets tighten. Joint-processing pipelines and the simulations[6] needed to validate them are essential, and without them, systematics introduced in combination are difficult and costly to untangle (i.e., could be misread as real physical signals) after they've entered published results.
- Build on the interagency coordination that already occurs by giving it standing and resourcing to commission joint deliverables. Precedent exists: TCAN [62] was established jointly by NSF/AST and NASA/APD in direct response to Astro2010's finding that no mechanism supported sustained multi-institutional theory collaborations. Its trajectory is also instructive: the joint solicitation lapsed while NASA continued the program alone. Mechanisms of this kind work, but without durable standing and resourcing they depend on the continued attention of individuals in both agencies.
- Fund leadership with an explicit cross-survey mandate. Every capability in Section 3 is currently additional to someone's actual job; work that is nobody's primary responsibility advances at the pace of spare time.

---

[4] like the LSST Discovery Alliance's Catalyst program
[5] Roman's default photometric redshifts, for example, are derived from Roman data alone rather than jointly utilizing optical photometry
[6] NASA's OpenUniverse [41] - a modest, sustained NASA-funded effort supported by DOE supercomputing that produces cross-survey simulation infrastructure - is an example of this kind of investment in simulations. OpenUniverse is scoped to simulation tools and infrastructure; it does not address joint processing of real survey data or fund the production of simulations for any survey.

- Build policy tools for collaboration, such as specific guidelines or templates for Memorandums of Understanding (MOUs) and documented procedures for enacting these between collaborations and agencies.

### 4.2 Observatories/Operators

Several of the building blocks this document describes exist thanks to the observatories, missions, and their operating organizations building them ahead of any mandate. The steps below extend that.

- Support community efforts toward *public*, jointly analyzable deep fields. One or a few small fields (e.g., EDF-S, COSMOS), carefully selected with the missions and their collaborations, would serve as pathfinders for unrestricted multi-survey analysis, with lessons that will extend naturally across the archives as the ecosystem matures[7].
- Share what's already working. Exchanging information across mission science collaborations about tools and approaches that have proven out keeps effort from being needlessly duplicated across surveys.
- Coordinate the observations. Combined datasets are most valuable when obtained contemporaneously -- especially for time-domain science [11] -- and observatories can both align observation planning where mission constraints allow and jointly articulate to their sponsoring agencies where a formal coordination mechanism is still needed.
- Streamline approval for coordinated triggers. Fully exploiting jointly-triggered transient observations -- including targeted follow-up of limited-lifetime events or solar-system objects outside either survey's baseline plan -- is achievable in part through existing guest-observer and target-of-opportunity processes. NASA's ACROSS initiative [68] is beginning to build this layer for its space observatories; extending such coordination across agency boundaries, to include ground-based facilities such as Rubin and the ground-based surveys that follow, remains the open step.

### 4.3 Universities, Science Centers, and National Computing Facilities

- Invest in modern computing infrastructure for joint processing. Joint processing at this scale will require substantial, scalable compute -- including GPUs and other accelerators -- beyond what any single survey's existing processing stack supports, funded on a model that accommodates regular hardware refresh rather than the build-once profile of traditional mission and observatory budgets.
- Fund the people, not just the code. Porting joint-processing codes to modern architectures is skilled software work; universities and computing centers can create the positions that make it someone's actual job rather than everyone's side project.
- Support cloud-resident copies of pixel data. Despite egress costs, cloud residency shifts the constraint from competition for finite, survey-dedicated hardware to questions of access,

---

[7] Archive interoperability and cross-survey information sharing matter: transparent, cross-accessible datasets and coordinated scheduling help success

API design, and sustainable funding. Universities and computing centers are natural partners in maintaining these data and building joint-processing capabilities around them.
- Provide a durable institutional home for researchers at the surveys-AI intersection. These researchers currently have no natural department or funding line to sit within (Section 3.4); universities are where that home gets built -- through joint appointments, cross-departmental positions, and fellowship lines that make interdisciplinary work a career path rather than a trap; but experience shows that such efforts, when supported as short-term projects, dissolve when the funding ends. What is needed are standing structures that sustain the space between disciplines, and across funding cycles.

## 4.4 The Research Community

- Share data and coordinate through existing platforms instead of building new ones[8]. Every team is naturally inclined to build a system it controls -- but proliferating one-off data-sharing tools recreates the same fragmentation this call to action is meant to solve. Contributing to, and coordinating through, platforms the community has already built is faster and cheaper than starting over each time.
- Continue developing and open-sourcing joint-processing tools and cross-survey algorithms, even ahead of formal funding -- as parts of the community have already begun to do.
- Host workshops and challenges that bring cross-survey efforts together[9]. Dedicated, focused convenings of this kind are worth applying to each pillar above: joint pixel-level processing, AI-substrate benchmark design, and standardized access protocols alike.
- Update graduate and postdoctoral training to include the cross-survey, AI-adjacent skills this work requires -- the interdisciplinary training gap this call to action identifies starts, in part, in how researchers are trained.
- Engage when funders and institutions convene structured community input; and volunteer perspectives proactively.

## 4.5 Philanthropy and Community Organizations

- Philanthropies can play a catalytic role in the development of cross-survey capabilities: in research groups, in ecosystem development and support, and in the people and career pathways at the intersection.
- Philanthropies can also support narrative and community building around cross-survey science, as a key lever for scientific progress and an understated aspect of science funding policy.
- Community organizations can also serve as standard-setting and convening bodies, in addition to federal agencies. The IVOA has long played this role for astronomical data interoperability, and professional societies and consortia can extend it to cross-survey standards for joint products, benchmarks, and access protocols.

[8] For example, NASA's Fornax science cloud [65], the Rubin Science Platform [66], and the Roman Research Nexus [67], among others; see also Section 3.3.
[9] Such as the June 2026 Broker Workshop that showed this model works: bringing the community together resolved design questions for cross-matched alert streams.

## 5. Conclusions

The coming years present an extraordinary moment of opportunity for survey astronomy. Combining observations across facilities is not new; the field has been doing it for four decades. What is new is that flagship wide-field surveys will observe the same sky concurrently, each producing petabyte-scale datasets, and that their greatest discoveries will be revealed across the seams. A public investment exceeding $6 billion has built the telescopes; but what remains unbuilt is the layer between them. We believe the full scientific return of these surveys requires treating that layer -- joint processing and validation, the AI-ready data substrate, interoperable access, and the people who sustain them -- as infrastructure in its own right, as essential as the telescopes themselves. This can likely be achieved at a small fraction of the cost of building and operating the observatories. The window in which foundational choices can still be aligned -- inexpensively, by design from the start rather than by costly reconciliation later -- is open now. It will not stay open, and it will not return. The cross-survey decade has begun; whether its science arrives in full is the choice this community, its institutions, and its supporters make in the next few years.

# Appendix A: The Cross-Survey Science Landscape in Full

This appendix expands the science cases summarized in Section 2.

## A.1 COSMOLOGY

**Weak lensing & dark energy:** Rubin delivers weak-lensing shapes over ~18,000 sq deg, but its photometric redshifts suffer catastrophic outliers exactly where it matters [39, 47], and ground-based blending biases the shear. Roman's near-infrared photometry improves the photo-z problem, and its stable space-based PSF offers a route to calibrate Rubin's systematics at the pixel level; jointly, the two tighten the equation-of-state constraints beyond what either achieves alone and cross-check systematics neither can diagnose alone. This is a flagship case that is *impossible from the planned public products*: it requires joint pixel-level processing to calibrate shear and selection. (Forced photometry on Rubin using Roman achieves some of it, but leaves out all of Rubin's deep-imaging contribution to an object's shape.) The same processing then unlocks deblended photometry, photo-z's, low-surface-brightness/dwarf-galaxy science, and better supernova-systematics control across optical+near-IR.

**Photometric redshifts & SED fitting:** Adding Roman's near-IR photometry and higher-resolution imaging to Rubin's six optical bands improves photo-z estimation -- and therefore cosmological constraints -- while sharpening DCR-based SED estimates and star-galaxy separation [9]. Beyond photo-z, the broader wavelength baseline improves estimates of stellar mass, star-formation rate, and dust attenuation by helping to reduce age-dust-metallicity degeneracies.

**Supernova (Ia) systematics & transient hosts:** Dust is a major concern for SN Ia systematics [48]; well-characterized transient light curves and host-galaxy models across optical+near-IR would reduce it. Coherent multi-band light curves arise for transients in overlapping time-domain fields (though not necessarily optimally), and the host-galaxy modeling comes essentially free from the joint data products above. This feeds back into cosmology via the SN Ia mass step and possible brightness-dust correlations. Sample studies of transient and variable sources host galaxies are also of interest for multimessenger-source progenitors such as dual AGN and candidate massive black hole binaries.

## A.2 TIME-DOMAIN & TRANSIENTS

**A jointly triggered transient stream**: Rubin discovers the transient and rapidly triggers Roman observations for selected events requiring high-resolution near-IR characterization -- for example, dust-obscured supernovae, nuclear transients, and events in crowded or high-extinction regions. Such a cross-mission triggering capability is in neither survey's baseline. Joint scene modeling can then extract light curves by fitting time-variable sources directly, without first constructing pixel-space difference templates, reducing startup delays when adequate templates do not yet exist [23, 24].

The single-survey time-domain foundations are rapidly maturing on both sides. On the Roman side, RAPID[10] (Roman Alerts Promptly from Image Differencing; [10]) -- a Caltech/IPAC Project Infrastructure Team -- will deliver prompt image differencing, a public alert stream of transient and variable candidates, and forced-photometry services, with machine-learning classification feeding the community brokers. On the Rubin side, the LSST alert stream and community brokers are already operating and beginning to ingest multi-messenger streams. Each survey's transient infrastructure is being built to do precisely what it was chartered to do -- and each charter is, deliberately, single-survey. What remains in no one's scope is the joint layer: brokers that merge the Rubin and Roman streams and enrich each alert with joint metadata -- best-available photo-z, cross-survey light-curve history, host context -- triggering jointly on events that require both surveys, and modeling the shared sky as one scene.

[10] https://rapid.ipac.caltech.edu/

**Transient host identification, characterization & classification:** Roman's high-resolution imaging characterizes a transient's host-galaxy morphology and properties in detail, and localizes the transient's position to high accuracy along with its local environment -- critical for classifying some events, including exotic ones like off-nuclear tidal disruption events. Additionally, joint searches can identify events that either survey alone would miss or classify ambiguously. Kilonovae and gravitational-wave counterparts show a hallmark rapid reddening from bound-bound opacity in freshly synthesized heavy elements; serendipitous kilonova searches and triggered GW-counterpart searches both benefit sharply from searching Rubin and Roman data jointly, rather than in series - and the same joint catalogs underpin statistical standard-siren cosmology, where redshift information comes from the survey itself [45, 46].

The highest-redshift supernovae can only be disentangled from low-redshift contaminants using deep optical non-detections from Rubin combined with bright Roman detections -- neither survey's data alone resolves the ambiguity. And stellar mergers, being intrinsically self-obscured, are best identified from the shape of the full Rubin+Roman spectral energy distribution rather than from either survey's photometry alone. At the limit, the survey combination does not improve a measurement -- it is what makes the discovery possible at all.

**Transients Roman sees that Rubin doesn't:** dust-obscured tidal disruption events, or AGN flares reprocessed at longer wavelengths that may be associated with mergers of stellar-mass binary black holes; in both cases the relative rates of obscured vs. unobscured transients inform the rates of possible multimessenger sources in the LIGO/Virgo/KAGRA and LISA bands.

**Multi-survey, multi-messenger time domain**: the LSST community brokers are beginning to incorporate Roman transients, and merging/matching the ZTF and LSST streams is straightforward as LSST settles into operations. Joint consideration extends to multi-wavelength (ALMA, Fermi, Swift, VLA) and multi-messenger streams -- the LVK gravitational-wave stream is being incorporated into brokers, and SNEWS (neutrinos) is restarting with new funding. Galactic-plane/bulge variables and transients will see renewed challenges with Roman, where joint scene modeling is specifically needed.

**Microlensing**: microlensing adds a further case, as planetary and free-floating-planet signatures appear as short deviations on longer events, so the science depends on identifying anomalies while they are ongoing. Roman's bulge time-domain survey will operate alongside existing ground-based alert infrastructure, and integrating these streams, rather than reconciling catalogues afterwards, is what makes coordinated response possible on the timescales the anomalies allow.

## A.3 GALAXIES, AGN & BLACK HOLES

**Galaxy formation & environment:** The deep joint optical+near-IR and deblending gains open new avenues for the most interesting classes of objects -- dwarf and low-surface-brightness galaxies -- where identifying them and measuring their masses via weak lensing is challenging for either survey alone. This comes essentially free from the data products needed for weak lensing. Joint pixel analysis also lets us probe more deeply and accurately for low-surface-brightness objects generally, from distant AGN and supernova light echoes to faint galaxies (including low-mass members of the Local Group).

**Supermassive black holes & AGN:** Building accurate, scalable cross-project deblending and photometry will allow combination not just of Rubin, Roman, and Euclid but also HSC, Pan-STARRS, and CFHTLS. (The Euclid deep fields -- 53 deg² -- reach 5σ of 27.2 AB mag in a single red optical band and ~26 mag in y, J, H, roughly Roman's depth.) The resolution of the space datasets allows superior separation of AGN light from starlight, enabling fundamental questions about the origin of supermassive black holes to be answered.

**Dust maps for galaxies:** Roman's deep high-resolution NIR imaging (in contrast to Euclid's NISP), together with Rubin's optical filters, allows the creation of dust maps for the majority of all galaxies, pushing dust/ICM studies to higher redshift and complete samples (and enabling the dust-corrected photometry noted under Cosmology) [44].

Determining the true astrophysical "sky" (extended Galactic/Local-Group emission, Galactic cirrus, intra-cluster light) as you push to low surface brightness from ground and space requires simultaneous mapping in space and wavelength -- a joint pixel-level analysis needing coordinated image calibration. Full scene modeling including differential chromatic refraction also yields finer SED information than the filters alone.

## A.4 STARS & THE MILKY WAY

**Stellar variability & Galactic structure:** Time-series photometry across the whole optical+NIR range, with overlapping passbands, lets a wealth of stellar variability be fully characterized -- e.g., the period-color-luminosity relation for thousands of RR Lyrae will enable studies of Galactic structure through previously dust-obscured regions of the Galaxy. This is also vital for excluding known variable sources from transient classification. Roman's high resolution and precision astrometry deblend Rubin images in crowded regions, enabling more precise cross-matching and much-improved color photometry for billions of stars; Rubin's timeseries astrometry helps constrain proper motions, and multi-epoch observations with Roman can greatly improve proper motion precisions especially for sources with high optical extinction beyond the distance of Galactic center.

Away from the Galactic plane, at high latitude, this works cleanly: independent Roman and Rubin point-source catalogs can be reasonably cross-matched and combined. In the plane itself -- where most of the Galaxy's stars and dust actually are -- this breaks down. Crowding makes catalog cross-matching ambiguous and limits the quality of Rubin's optical photometry outright; the Roman catalogs need to be brought back to the Rubin pixels for sensible joint photometry across the majority of stars Roman detects there. This matters acutely for the Roman Galactic Plane Survey [11], which observes primarily in F129 and redward -- a limited lever arm on its own for constraining stellar types. Rubin's photometry extends that lever arm smoothly blueward, but only once a consistent, jointly-derived photometric system exists across both surveys.

A related computational challenge sits beneath the joint photometric system itself: fully simultaneous, regularized 3D dust inversion across sight lines remains practical today at the ~$10^7$-star scale, while maps at the ~$10^8$–$10^9$-star scale [e.g., 40] have so far relied on per-sightline inference with spatial priors. Scaling the fully simultaneous technique to billion-object catalogs, while properly propagating distance uncertainties, remains an open computational problem in its own right.

Combined, this massively extends the map of Galactic structure and kinematics given Gaia's incompleteness in crowded regions -- and could even allow exoplanet-transit searches in Rubin data that would otherwise be too crowded.

**Low-mass stars:** These are u-g-r-i dropouts in Rubin data; joint Rubin and Roman can confirm their nature and characterize their SEDs while retaining time-domain characterization. They are among the hardest targets to observe, with largely unknown properties at the low-temperature end (see [38]). Extending Rubin to IR also penetrates interstellar dust in the Galactic plane and minimizes crowding confusion through improved angular resolution.

## A.5 SOLAR SYSTEM

**Solar-system & near-Earth objects:** Joint parallax of moving objects -- leveraging Roman's wide field and non-destructive reads with Rubin's cadence -- delivers astrometry beyond any ground-based pair, with direct application to detecting and characterizing potentially Earth-impacting asteroids and to probing the outer solar system.

Shift-and-stack searches for trans-Neptunian objects have been demonstrated within single optical surveys (including Rubin) but not cross-filter or cross-survey; coupling Roman's sensitivity and resolution with Rubin's high intra-night cadence in Deep-Drilling Fields could detect and characterize such populations beyond either survey or a catalog combination.

## 6. References

[1] Rubin-LSST survey start announcement: https://rubinobservatory.org/news/action-rubin-lsst-begins
[2] ROTAC Report: https://roman.gsfc.nasa.gov/science/ccs/ROTAC-Report-20250424-v1.pdf
[3] GOODS: https://www.stsci.edu/science/goods/DataProducts/
[4] Gezari, S., Bentz, M., et al., R2-D2: Roman and Rubin -- From Data to Discovery, preprint arXiv:2202.12311 (2022)
[5] Guy, L. P., et al., Rubin-Euclid Derived Data Products: Initial Recommendations, preprint arXiv:2201.03862 (2022)
[6] Rhodes, J., et al., Scientific Synergy Between LSST and Euclid, ApJS, 223, 21,(2017)
[7] Bachelet, E., et al., Euclid-Roman joint microlensing survey: early mass measurement, free floating planets and exomoons, A&A, 664, 136, (2022) [https://arxiv.org/abs/2202.09475]
[8] Chary, R., et al., Joint Survey Processing of LSST, Euclid and WFIRST: Enabling a broad array of astrophysics and cosmology through pixel level combinations of datasets, preprint arXiv:1910.01259 (2019) [https://arxiv.org/abs/1910.01259]
[9] Bianco, F. B., et al., Maximizing the scientific return of Roman and Rubin with a joint wide-sky observing strategy, preprint arXiv:2402.02378 (2024)
[10] Gandhi, K., et al., "Identifying Gems from Roman RAPIDly," Publications of the Astronomical Society of the Pacific, 138 (2026) [https://iopscience.iop.org/article/10.1088/1538-3873/ae7904]
[11] Kruszyńska, K., et al., Synergies between Roman Galactic Plane Survey and other major surveys, preprint arXiv:2406.14767 (2024)
[12] York, D. G., Adelman, J., Anderson, J. E., Jr., et al., "The Sloan Digital Sky Survey: technical summary," Astron. J. 120(3), 1579-1587 (2000). [arXiv:astro-ph/0006396]
[13] Ivezić, Ž., Kahn, S. M., Tyson, J. A., et al., "LSST: from science drivers to reference design and anticipated data products," Astrophys. J. 873(2), 111 (2019). [arXiv:0805.2366]
[14] Laureijs, R., Amiaux, J., Arduini, S., et al. 2011, Euclid Definition Study Report, ESA/SRE(2011)12, arXiv:1110.3193
[15] Spergel, D., Gehrels, N., Baltay, C., et al. 2015, Wide-Field InfrarRed Survey Telescope-Astrophysics Focused Telescope Assets WFIRST-AFTA 2015 Report, arXiv:1503.03757
[16] Chary, R., et al. 2020, Joint Survey Processing of Euclid, Rubin and Roman: Final Report, arXiv:2008.10663
[17] Alonso, D., et al. 2021, Combining information from multiple cosmological surveys: inference and modeling challenges, arXiv:2103.05320
[19] Blanton, M. R., et al. 2023, The Future of Astronomical Data Infrastructure: Meeting Report, arXiv:2311.04272.
[20] Borgman, C. L. & Wofford, M. F. 2021, From Data Processes to Data Products: Knowledge Infrastructures in Astronomy, arXiv:2109.01707.
[21] Melchior, P., Moolekamp, F., Jerdee, M., et al., "SCARLET: source separation in multi-band images by constrained matrix factorization," Astron. Comput. 24, 129-142 (2018). [arXiv:1802.10157]
[22] Longley, E. P., et al. 2023, "A unified catalogue-level reanalysis of stage-III cosmic shear surveys," Monthly Notices of the Royal Astronomical Society, 520, 5016-5041. doi:10.1093/mnras/stad246
[23] Ward, C., Melchior, P., Sampson, M. L., et al., "Disentangling transients and their host galaxies with scarlet2: a framework to forward model multi-epoch imaging," Astron. Comput. 51, 100930 (2025). [arXiv:2409.15427]

[24] Mendoza, I., Hansen, D., Liu, R., et al. (LSST Dark Energy Science Collaboration), "Simulation-based inference for probabilistic galaxy detection and deblending," arXiv:2601.03422 (2026)
[25] Troxel, M. A., et al. 2023, "A joint Roman Space Telescope and Rubin Observatory synthetic wide-field imaging survey," Monthly Notices of the Royal Astronomical Society, 522, 2801-2820. doi:10.1093/mnras/stad664
[26] Euclid Collaboration: Y. Mellier, Abdurro'uf, J. A. Acevedo Barroso, et al., "Euclid. I. Overview of the Euclid mission," preprint, arXiv:2405.13491 (2024)
[27] Ferguson, A., LaFleur, M., Ruthotto, L., et al., "The future of artificial intelligence and the mathematical and physical sciences (AI+MPS)," arXiv:2509.02661 (2025).
[28] The Multimodal Universe Collaboration, 'The Multimodal Universe: enabling large-scale machine learning with 100 TB of astronomical scientific data,' Proc. NeurIPS Datasets and Benchmarks (2024). [arXiv:2412.02527]
[29] Deng, J., Dong, W., Socher, R., Li, L.-J., Li, K. and Fei-Fei, L., "ImageNet: a large-scale hierarchical image database," Proc. IEEE Conf. Computer Vision and Pattern Recognition (CVPR), 248-255 (2009)
[30] Russakovsky, O., Deng, J., Su, H., et al., "ImageNet large scale visual recognition challenge," Int. J. Comput. Vis. 115(3), 211-252 (2015). [arXiv:1409.0575]
[31] Polymathic AI: https://polymathic-ai.org/
[32] SkAI: https://skai-institute.org/
[33] CosmicAI: https://cosmicai.org/
[34] https://datalabs.esa.int/
[35] Caplar, N., Beebe, W., Branton, D., et al., "Using LSDB to enable large-scale catalog distribution, cross-matching, and analytics," arXiv:2501.02103 (2025)
[36] LINCC Frameworks: https://lsstdiscoveryalliance.org/lsst-discovery-alliance-programs/lincc-frameworks/
[37] IVOA: https://www.ivoa.net/
[38] Easton J. Honaker et al., "Searching for Ultracool Dwarfs in Early LSST Data Products", The Astrophysical Journal in June 2026, DOI 10.3847/1538-4357/ae6ce9
[39] Mandelbaum, R., et al., "The LSST Dark Energy Science Collaboration (DESC) Science Requirements Document," preprint arXiv:1809.01669 (2018).
[40] Green, G. M., Schlafly, E. F., Zucker, C., Speagle, J. S., and Finkbeiner, D. P., "A 3D Dust Map Based on Gaia, Pan-STARRS 1 and 2MASS," preprint arXiv:1905.02734 (2019)
[41] The OpenUniverse Collaboration, "OpenUniverse2024: A shared, simulated view of the sky for Rubin and Roman," preprint arXiv:2501.05632 (2025)
[42] Nikutta, R., Fitzpatrick, M., Scott, A., and Weaver, B. A., "Data Lab - a community science platform," Astron. Comput. 33, 100411 (2020)
[43] Juneau et al. 2021, "Jupyter-Enabled Astrophysical Analysis Using Data-Proximate Computing Platforms," CiSE 23, 15 (2021)
[44] Ménard, B., Scranton, R., Fukugita, M. and Richards, G., "Measuring the galaxy-mass and galaxy-dust correlations through magnification and reddening," Mon. Not. R. Astron. Soc. 405(2), 1025-1039 (2010). [arXiv:0902.4240]
[45] Palmese, A., Bom, C. R., Mucesh, S. and Hartley, W. G., "A standard siren measurement of the Hubble constant using gravitational-wave events from the first three LIGO/Virgo observing runs and the DESI Legacy Survey," Astrophys. J. 943, 56 (2023). [arXiv:2111.06445]

[46] Palmese, A., deVicente, J., Pereira, M. E. S., et al., "A statistical standard siren measurement of the Hubble constant from the LIGO/Virgo gravitational wave compact object merger GW190814 and Dark Energy Survey galaxies," Astrophys. J. Lett. 900, L33 (2020). [arXiv:2006.14961]
[47] Newman, J. A. and Gruen, D., "Photometric redshifts for next-generation surveys," Annu. Rev. Astron. Astrophys. 60, 363-414 (2022). [arXiv:2206.13633]
[48] Rose, B. M., et al., "Synergies between Vera C. Rubin Observatory, Nancy Grace Roman Space Telescope, and Euclid Mission: constraining dark energy with Type Ia supernovae," arXiv:2104.01199 (2021)
[49] Norman, D., et al., "The growing importance of a tech-savvy astronomy and astrophysics workforce," Astro2020 APC white paper, arXiv:1907.07184 (2019)
[50] Parker, L., Lanusse, F., et al., "AstroCLIP: a cross-modal foundation model for galaxies," Mon. Not. R. Astron. Soc. 531(4), 4990-5011 (2024). [arXiv:2310.03024]
[51] Melchior, P., Joseph, R., Sanchez, J., MacCrann, N., and Gruen, D., “The challenge of blending in large sky surveys”, Nature Reviews Physics, 3 (10), 712-718 (2021)
[52] Joseph, R., Melchior, P., and Moolekamp, F., “Joint survey processing: combined resampling and convolution for galaxy modelling and deblending”, arXiv:2107.06984 (2021)
[53] Melchior, P., Ward, C., Remy, B., Wiemann, M., and Siegel, J., “scarlet2: Astronomical scene modeling in JAX”, The Journal of Open Source Software, 11, 120, 9646 (2026)
[54] Audenaert, J., Muthukrishna, D., Gregory, P. F., Hogg, D. W., and Villar, V. A., “Causal Foundation Models: Disentangling Physics from Instrument Properties”, arXiv:2507.05333 (2025)
[55] Romelli, E., et al., “Euclid Quick Data Release (Q1): From images to multiwavelength catalogues: the Euclid MERge Processing Function”, https://arxiv.org/pdf/2503.15305
[56] Harwit & Neal, 1986, "The great observatories for space astrophysics", https://ntrs.nasa.gov/citations/19860015241
[57] Córdova, F. A. (ed.), Multiwavelength Astrophysics, Cambridge University Press, Cambridge & New York, 1988, 400 pp. ISBN 0-521-36197-4
[58] Jumper et al. 2021, "Highly accurate protein structure prediction with AlphaFold," Nature 596, 583-589
[59] Offner, S. S. R., "AI Reaches for the Stars," Dædalus 155 (1), Winter/Spring 2026. https://www.amacad.org/publication/daedalus/ai-reaches-stars
[60] Scoville, N., Aussel, H., Brusa, M., et al., "The Cosmic Evolution Survey (COSMOS): Overview," ApJS 172, 1 (2007). doi:10.1086/516585
[61] Abbott, T. M. C., et al. "DES Y3 + KiDS-1000: Consistent Cosmology Combining Cosmic Shear Surveys." The Open Journal of Astrophysics, vol. 6, 2023, https://arxiv.org/abs/2305.17173
[62]https://www.nsf.gov/funding/opportunities/tcan-theoretical-computational-astrophysics-networks/504843/nsf13-512
[63] https://roman-hlis-cosmology.caltech.edu/
[64] Sanchez, J., Mendoza, I., Kirkby, D. P., Burchat, P. R., and LSST Dark Energy Science Collaboration, “Effects of overlapping sources on cosmic shear estimation: Statistical sensitivity and pixel-noise bias”, Journal of Cosmology and Astroparticle Physics, vol. 2021, no. 7, Art. no. 043, IOP, 2021. doi:10.1088/1475-7516/2021/07/043
[65] https://science.nasa.gov/astrophysics/programs/physics-of-the-cosmos/community/the-fornax-initiative/
[66] https://data.lsst.cloud/

[67] https://roman-docs.stsci.edu/data-handbook/roman-research-nexus
[68] https://science.data.nasa.gov/data-sites/across